# Large energy mode locking of an erbium-doped fiber laser with atomic layer graphene

H. Zhang[1], D. Y. Tang[1]*, L. M. Zhao[1], Q. L. Bao[2], K. P. Loh[2]

[1]School of Electrical and Electronic Engineering, Nanyang Technological University, Singapore 639798

[2]Department of Chemistry, National University of Singapore, Singapore 117543

*Corresponding author: edytang@ntu.edu.sg

We report on large energy pulse generation in an erbium-doped fiber laser passively mode-locked with atomic layer graphene. Stable mode locked pulses with single pulse energy up to 7.3 nJ and pulse width of 415 fs have been directly generated from the laser. Our results show that atomic layer graphene could be a promising saturable absorber for large energy mode locking.

OCIS codes: 060.5530, 060.2410, 140.3510, 190.5530.

High power ultrashort optical pulses have widespread applications in industrial material processing, medical treatment and scientific researches. Recently, it has been demonstrated that such optical pulses can also been directly generated by the passively mode locked fiber laser oscillators. By using the all-normal dispersion dissipative soliton shaping technique, Ruehl et al. have demonstrated 10 nJ pulse generation in an Erbium doped fiber (EDF) laser [1], and A. Chong et al have generated 20 nJ pulses from a Yb-doped fiber laser [2]. The results have shown the promise of using the mode locked fiber lasers to replace the bulk ultrashort pulse solid-state lasers for many practical applications. However, the reported high power mode locked fiber lasers have a drawback. In order to enable the high power mode locking operation, an artificial saturable absorber formed based on the light interference was used as the mode locker. Therefore, the lasers are environmentally unstable. In order to sidestep this drawback, a real passive mode locker, particularly single walled carbon nanotubes (SWCNTs) based saturable absorber [3-11], which can endure high optical power should be used. Very recently, Y. W. Song et al reported a technique of using single walled carbon nanotubes as mode locker for high-energy pulse formation [11]. By employing the evanescent field interaction of light with the nanotubes, it was shown that the high optical power induced damage of the nanotube could be avoided. Consequently, stable mode locked pulses with 1.2ps pulse width and 6.5 nJ pulse energy were generated in an erbium-doped fiber laser. In this letter we report on the large energy ultrashort pulse generation in an erbium-doped fiber laser passively mode locked with atomic layer graphene. We show that graphene can not only be used as a mode locker to mode lock fiber lasers, but also has high optical damage threshold.

Stable mode locked pulses with pulse energy as high as 7.3 nJ and pulse width of 415fs have been directly obtained from a dispersion-managed cavity fiber lasers.

Graphene is a single two-dimensional (2D) atomic layer of carbon atom arranged in a hexagonal lattice. An isolated graphene film is a zero bandgap semiconductor with a linear energy dispersion relation for both electrons and holes near Dirac point [12]. Saturable absorption in graphene is achieved due to the Pauli blocking of the electrons and holes for occupation of the energy levels in the conduction and valence bands that are resonant with the incident photons [13]. Recent advance in graphene research has shown that graphene saturable absorption has an ultrashort recovery time [13]. As graphene has a 2D structure, it has much smaller non-saturable loss and higher damage threshold. Therefore, it is expected that using graphene as a laser mode locker large energy ultrashort pulses could be generated.

Our fiber laser is schematically shown in Fig. 1. A piece of 5.0 m, 2880 ppm Erbium-doped fiber (EDF) with group velocity dispersion (GVD) of -32 (ps/nm)/km was used as the gain medium, and 24.2 m single mode fiber (SMF) with GVD 18 (ps/nm)/km was employed to compress the intra-cavity pulses and obtain ultrashort pulses. The net cavity fiber dispersion is estimated -0.3583 $ps^2$. A 30% fiber coupler was used to output the signal. The laser was pumped by a high power Fiber Raman Laser source (KPS-BT2-RFL-1480-60-FA) of wavelength 1480 nm, and the maximum pump power can be as high as 5 W. A polarization independent isolator was spliced in the cavity to force the unidirectional operation of the ring cavity, and an intra-cavity polarization controller (PC) was used to fine tune the linear cavity birefringence.

The graphene mode locker used in our experiment was made by transferring a piece of free standing graphene film onto a fiber pigtail by the mutual Van Der Waals forces. The few layers graphene thin film was synthesized on Ni/Si substrate by the chemical vapor deposition (CVD) method [13, 14]. As-produced graphene was stripped off the substrate through an oxidizing etching treatment in an aqueous iron (III) chloride (FeCl3) solution (~ 1M) to obtain a freestanding atomic layer graphene film. Fig. 2a shows a Raman spectrum of the graphene, from which the G band and 2D band are clearly resolved. The relative weak D band indicates very high crystallinity of our samples. Fig. 2b shows the Raman image plotted by the Raman peak of doped $SiO_2$, from which the fiber core area is located. And Fig. 2c shows the Raman image plotted by G band of graphene around core area, which demonstrates that few layers graphene is coated on the fiber core. We have experimentally measured the absorption modulation depth and saturation fluence of the graphene saturable absorber. They are ~**66.5%** and **0.71** MW/cm$^2$, respectively [13].

Self-starting mode locking of the laser occurred at the incident pump power of about 130 mW. The mode locking state could then be maintained to the maximum accessible pump intensity of ~ 3.5W. Fig.3 shows a typical mode locking state of the laser. Fig. 3a is the optical spectrum of the mode locked pulses. It is centered at 1576.3nm and has a 3dB bandwidth of ~10nm. Although the net cavity fiber dispersion is anomalous, no Kelly sidebands are observed on the spectrum. Fig. 3b is the measured autocorrelation trace of the mode locked pulses. It has a FWHM width of 590 fs. If a Gaussian-pulse profile is assumed, the pulse duration is 415fs. The time-bandwidth product (TBP) of the pulses is 0.518, indicating that the mode locked pulses are slightly chirped. Fig. 3c shows the measured oscilloscopes trace within nanosecond and millisecond (insert of Fig. 3c) time scale. In all of the pump power range, the laser always emitted single pulse, no pulse breaking or multiple pulse operation was detected as confirmed with a high speed oscilloscope together with the autocorrelation trace measurement. The pulse circulated in the cavity with the fundamental cavity repetition time of 146 ns. Gradually increasing the pump strength, except the spectral bandwidth became broader, the autocorrelation pulse profile and the spectral profile had always the same shapes. We also measured the radio-frequency (RF) spectrum of the mode locking state. Fig. 3d shows a measurement made at a span of 10 kHz and a resolution bandwidth of 10 Hz. The fundamental peak located at the cavity repetition rate of 6.84 MHz has a signal-to-noise ratio of 65 dB. The insert of Fig. 3d shows the wideband RF spectrum up to 100 MHz. The absence of spectral modulation in RF spectrum demonstrates that the laser operates well in the CW mode-locking regime.

Fig. 4 shows the single pulse energy change with the pump power. Staring from the laser mod locking threshold, the output power increased linearly with the pump power with a slope efficiency of 6.7%. The maximum achieved single pulse energy is as high as ~7.3 nJ. To the best of our knowledge, this is the highest pulse energy reported for ultrafast erbium-doped fiber mode locked with a real saturable absorber in cavity. At the maximum output power, the pulse width is ~415fs, which gives the maximum peak power of 17.6 kW.

To investigate the long-term stability of the atomic layer graphene mode locking, we monitored the mode locked laser operation for 140 hours. As shown in the insert of Fig. 3a, the 3 dB spectral bandwidths of the output pulses were kept in a consistent value with little fluctuations. Despite of the fact that the graphene was radiated under an optical fluency of ~52 mJ/cm$^2$, no obvious degradation on the mode locking performance was observed. Optical microscopy study on the atomic layer graphene film also confirmed that its morphology has kept intact.

In summary, we have demonstrated large energy mode locking of an erbium-doped fiber laser with atomic layer graphene as the saturable absorber. Stable mode locked pulses with single pulse energy as high as 7.5nJ and 415 fs pulse width have been generated. Our experimental results shown that atomic layer graphene could be a promising saturable absorber for high power laser mode locking.

K. P. Loh wishes to acknowledge funding support from NRF-CRP Graphene Related Materials and Devices (R-143-000-360-281).

**Figure captions:**

Fig.1: Schematic of the fiber laser. WDM: wavelength division multiplexer. EDF: erbium doped fiber.

Fig.2: Characterization of graphene thin film covering on the fiber core. (a) Raman spectra of the graphene film. (b) Raman image around the fiber core plotted by the intensity of the Raman peak of SiO2. The scale bar is 3 µm. (c) Raman images around the fiber core plotted by the intensity of G band of graphene. The scale bar is 3 µm. (d) Normalized power dependent absorption of atomic layer graphene. Insert: the UV-VIS-NIR absorption spectra of graphene.

Fig. 3: Pulse operation of the fiber laser. (a) Pulse spectra measured. Insert: long-term fluctuation of the FWHM. (b) Autocorrelation traces of the pulses. (c) An oscilloscope trace of the single pulse emission. Insert: pulse train of CW mode-locking in millisecond time scale. (d) The fundamental radio-frequency (RF) spectrum of the laser output. Insert: wideband RF spectrum up to 100 MHz.

Fig. 4: the single pulse energy in respect to the pump power.

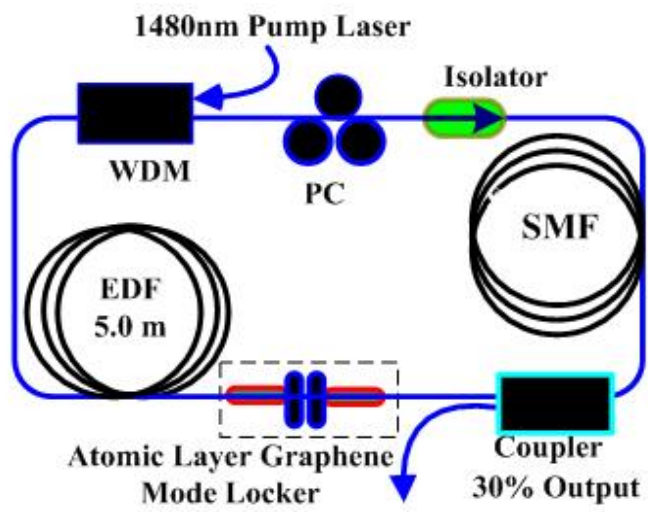

Fig.1: H. Zhang et. al.

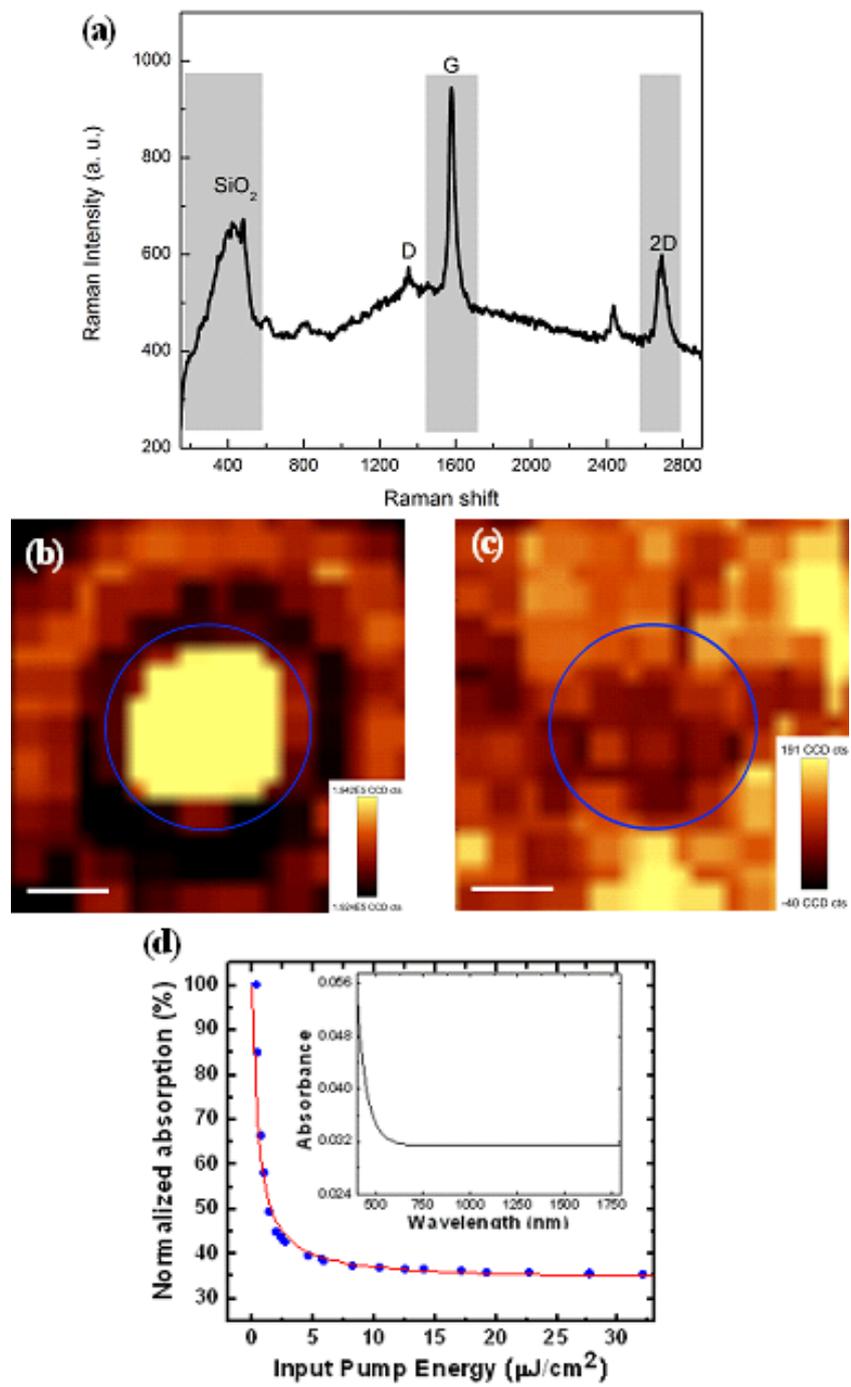

Fig.2: H. Zhang et. al.

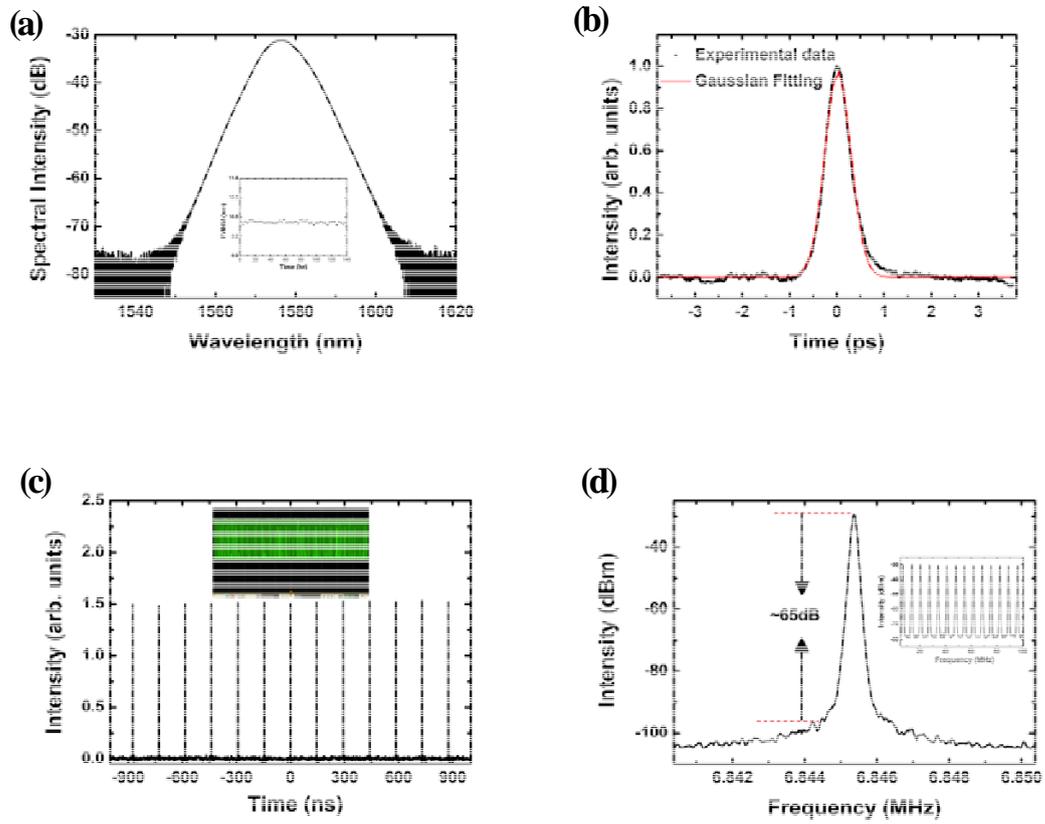

Fig.3: H. Zhang et. al.

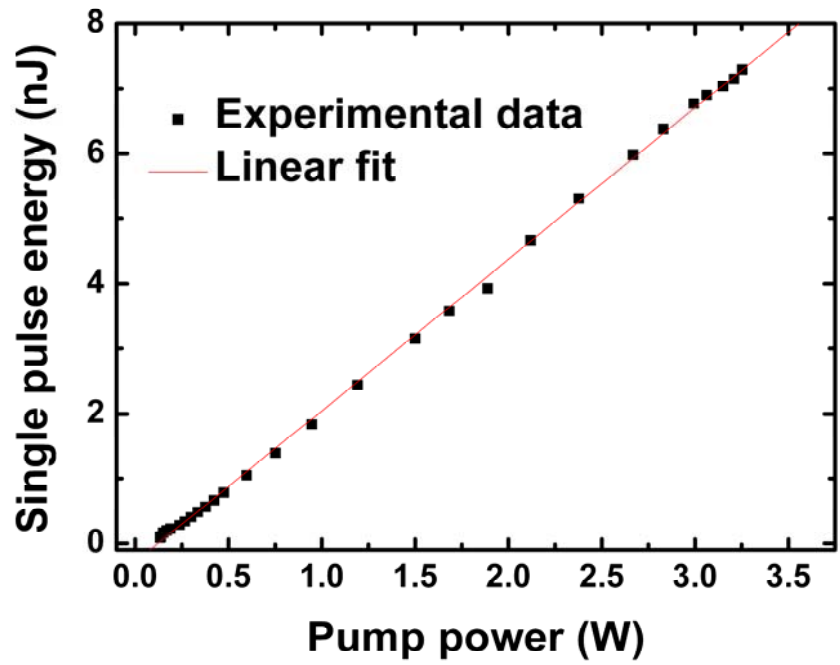

Fig.4: H. Zhang et. al.